\documentclass[conference,10pt,twoside,twocolumn]{IEEEtran}

\usepackage{cite}
\usepackage[T1]{fontenc}
\usepackage{graphicx}
\usepackage{amssymb}
\usepackage{amsmath}
\usepackage{amsthm}
\usepackage{subfigure}
\usepackage{booktabs} 
\usepackage{multirow}
\usepackage{microtype}
\usepackage{balance}
\usepackage{xcolor}
\usepackage{xfrac}    
\usepackage{algorithm}
\usepackage{setspace}

\usepackage{algorithmicx}
\usepackage{algpseudocode}
\algrenewcommand\algorithmicindent{0.7em}%
\usepackage{xpatch}

\usepackage{hyperref}

\usepackage{amssymb}
\usepackage{amsfonts}
\usepackage{mathrsfs}
\usepackage{xspace}
\usepackage{bm}
\usepackage{upgreek}

\newcommand{\safemath}[2]{\newcommand{#1}{\ensuremath{#2}\xspace}}

\safemath{\bma}{\mathbf{a}}
\safemath{\bmb}{\mathbf{b}}
\safemath{\bmc}{\mathbf{c}}
\safemath{\bmd}{\mathbf{d}}
\safemath{\bme}{\mathbf{e}}
\safemath{\bmf}{\mathbf{f}}
\safemath{\bmg}{\mathbf{g}}
\safemath{\bmh}{\mathbf{h}}
\safemath{\bmi}{\mathbf{i}}
\safemath{\bmj}{\mathbf{j}}
\safemath{\bmk}{\mathbf{k}}
\safemath{\bml}{\mathbf{l}}
\safemath{\bmm}{\mathbf{m}}
\safemath{\bmn}{\mathbf{n}}
\safemath{\bmo}{\mathbf{o}}
\safemath{\bmp}{\mathbf{p}}
\safemath{\bmq}{\mathbf{q}}
\safemath{\bmr}{\mathbf{r}}
\safemath{\bms}{\mathbf{s}}
\safemath{\bmt}{\mathbf{t}}
\safemath{\bmu}{\mathbf{u}}
\safemath{\bmv}{\mathbf{v}}
\safemath{\bmw}{\mathbf{w}}
\safemath{\bmx}{\mathbf{x}}
\safemath{\bmy}{\mathbf{y}}
\safemath{\bmz}{\mathbf{z}}
\safemath{\bmzero}{\mathbf{0}}
\safemath{\bmone}{\mathbf{1}}

\bmdefine{\biad}{a}
\bmdefine{\bibd}{b}
\bmdefine{\bicd}{c}
\bmdefine{\bidd}{d}
\bmdefine{\bied}{e}
\bmdefine{\bifd}{f}
\bmdefine{\bigd}{g}
\bmdefine{\bihd}{h}
\bmdefine{\biid}{i}
\bmdefine{\bijd}{j}
\bmdefine{\bikd}{k}
\bmdefine{\bild}{l}
\bmdefine{\bimd}{m}
\bmdefine{\bind}{n}
\bmdefine{\biod}{o}
\bmdefine{\bipd}{p}
\bmdefine{\biqd}{q}
\bmdefine{\bird}{r}
\bmdefine{\bisd}{s}
\bmdefine{\bitd}{t}
\bmdefine{\biud}{u}
\bmdefine{\bivd}{v}
\bmdefine{\biwd}{w}
\bmdefine{\bixd}{x}
\bmdefine{\biyd}{y}
\bmdefine{\bizd}{z}

\bmdefine{\bixid}{\xi}
\bmdefine{\bilambdad}{\lambda}
\bmdefine{\bimud}{\mu}
\bmdefine{\bithetad}{\theta}
\bmdefine{\biphid}{\phi}
\bmdefine{\bideltad}{\delta}

\safemath{\bmia}{\biad}
\safemath{\bmib}{\bibd}
\safemath{\bmic}{\bicd}
\safemath{\bmid}{\bidd}
\safemath{\bmie}{\bied}
\safemath{\bmif}{\bifd}
\safemath{\bmig}{\bigd}
\safemath{\bmih}{\bihd}
\safemath{\bmii}{\biid}
\safemath{\bmij}{\bijd}
\safemath{\bmik}{\bikd}
\safemath{\bmil}{\bild}
\safemath{\bmim}{\bimd}
\safemath{\bmin}{\bind}
\safemath{\bmio}{\biod}
\safemath{\bmip}{\bipd}
\safemath{\bmiq}{\biqd}
\safemath{\bmir}{\bird}
\safemath{\bmis}{\bisd}
\safemath{\bmit}{\bitd}
\safemath{\bmiu}{\biud}
\safemath{\bmiv}{\bivd}
\safemath{\bmiw}{\biwd}
\safemath{\bmix}{\bixd}
\safemath{\bmiy}{\biyd}
\safemath{\bmiz}{\bizd}

\safemath{\bmxi}{\bixid}
\safemath{\bmlambda}{\bilambdad}
\safemath{\bmmu}{\bimud}
\safemath{\bmtheta}{\bithetad}
\safemath{\bmphi}{\biphid}
\safemath{\bmdelta}{\bideltad}

\safemath{\bA}{\mathbf{A}}
\safemath{\bB}{\mathbf{B}}
\safemath{\bC}{\mathbf{C}}
\safemath{\bD}{\mathbf{D}}
\safemath{\bE}{\mathbf{E}}
\safemath{\bF}{\mathbf{F}}
\safemath{\bG}{\mathbf{G}}
\safemath{\bH}{\mathbf{H}}
\safemath{\bI}{\mathbf{I}}
\safemath{\bJ}{\mathbf{J}}
\safemath{\bK}{\mathbf{K}}
\safemath{\bL}{\mathbf{L}}
\safemath{\bM}{\mathbf{M}}
\safemath{\bN}{\mathbf{N}}
\safemath{\bO}{\mathbf{O}}
\safemath{\bP}{\mathbf{P}}
\safemath{\bQ}{\mathbf{Q}}
\safemath{\bR}{\mathbf{R}}
\safemath{\bS}{\mathbf{S}}
\safemath{\bT}{\mathbf{T}}
\safemath{\bU}{\mathbf{U}}
\safemath{\bV}{\mathbf{V}}
\safemath{\bW}{\mathbf{W}}
\safemath{\bX}{\mathbf{X}}
\safemath{\bY}{\mathbf{Y}}
\safemath{\bZ}{\mathbf{Z}}

\safemath{\bZero}{\mathbf{0}}
\safemath{\bOne}{\mathbf{1}}
\safemath{\bDelta}{\mathbf{\Delta}}
\safemath{\bLambda}{\mathbf{\UpLambda}}
\safemath{\bPhi}{\mathbf{\Upphi}}
\safemath{\bSigma}{\mathbf{\Upsigma}}
\safemath{\bOmega}{\mathbf{\Upomega}}
\safemath{\bTheta}{\mathbf{\Uptheta}}

\bmdefine{\biAd}{A}
\bmdefine{\biBd}{B}
\bmdefine{\biCd}{C}
\bmdefine{\biDd}{D}
\bmdefine{\biEd}{E}
\bmdefine{\biFd}{F}
\bmdefine{\biGd}{G}
\bmdefine{\biHd}{H}
\bmdefine{\biId}{I}
\bmdefine{\biJd}{J}
\bmdefine{\biKd}{K}
\bmdefine{\biLd}{L}
\bmdefine{\biMd}{M}
\bmdefine{\biOd}{N}
\bmdefine{\biPd}{O}
\bmdefine{\biQd}{P}
\bmdefine{\biRd}{R}
\bmdefine{\biSd}{S}
\bmdefine{\biTd}{T}
\bmdefine{\biUd}{U}
\bmdefine{\biVd}{V}
\bmdefine{\biWd}{W}
\bmdefine{\biXd}{X}
\bmdefine{\biYd}{Y}
\bmdefine{\biZd}{Z}

\bmdefine{\biDelta}{\Delta}
\bmdefine{\biLambda}{\Lambda}
\bmdefine{\biPhi}{\Phi}
\bmdefine{\biSigma}{\Sigma}
\bmdefine{\biOmega}{\Omega}
\bmdefine{\biTheta}{\Theta}

\safemath{\bimA}{\biAd}
\safemath{\bimB}{\biBd}
\safemath{\bimC}{\biCd}
\safemath{\bimD}{\biDd}
\safemath{\bimE}{\biEd}
\safemath{\bimF}{\biFd}
\safemath{\bimG}{\biGd}
\safemath{\bimH}{\biHd}
\safemath{\bimI}{\biId}
\safemath{\bimJ}{\biJd}
\safemath{\bimK}{\biKd}
\safemath{\bimL}{\biLd}
\safemath{\bimM}{\biMd}
\safemath{\bimN}{\biNd}
\safemath{\bimO}{\biOd}
\safemath{\bimP}{\biPd}
\safemath{\bimQ}{\biQd}
\safemath{\bimR}{\biRd}
\safemath{\bimS}{\biSd}
\safemath{\bimT}{\biTd}
\safemath{\bimU}{\biUd}
\safemath{\bimV}{\biVd}
\safemath{\bimW}{\biWd}
\safemath{\bimX}{\biXd}
\safemath{\bimY}{\biYd}
\safemath{\bimZ}{\biZd}

\safemath{\bimDelta}{\biDelta}
\safemath{\bimLambda}{\biLambda}
\safemath{\bimPhi}{\biPhi}
\safemath{\bimSigma}{\biSigma}
\safemath{\bimOmega}{\biOmega}
\safemath{\bimTheta}{\biTheta}

\safemath{\setA}{\mathcal{A}}
\safemath{\setB}{\mathcal{B}}
\safemath{\setC}{\mathcal{C}}
\safemath{\setD}{\mathcal{D}}
\safemath{\setE}{\mathcal{E}}
\safemath{\setF}{\mathcal{F}}
\safemath{\setG}{\mathcal{G}}
\safemath{\setH}{\mathcal{H}}
\safemath{\setI}{\mathcal{I}}
\safemath{\setJ}{\mathcal{J}}
\safemath{\setK}{\mathcal{K}}
\safemath{\setL}{\mathcal{L}}
\safemath{\setM}{\mathcal{M}}
\safemath{\setN}{\mathcal{N}}
\safemath{\setO}{\mathcal{O}}
\safemath{\setP}{\mathcal{P}}
\safemath{\setQ}{\mathcal{Q}}
\safemath{\setR}{\mathcal{R}}
\safemath{\setS}{\mathcal{S}}
\safemath{\setT}{\mathcal{T}}
\safemath{\setU}{\mathcal{U}}
\safemath{\setV}{\mathcal{V}}
\safemath{\setW}{\mathcal{W}}
\safemath{\setX}{\mathcal{X}}
\safemath{\setY}{\mathcal{Y}}
\safemath{\setZ}{\mathcal{Z}}
\safemath{\emptySet}{\varnothing}

\safemath{\colA}{\mathscr{A}}
\safemath{\colB}{\mathscr{B}}
\safemath{\colC}{\mathscr{C}}
\safemath{\colD}{\mathscr{D}}
\safemath{\colE}{\mathscr{E}}
\safemath{\colF}{\mathscr{F}}
\safemath{\colG}{\mathscr{G}}
\safemath{\colH}{\mathscr{H}}
\safemath{\colI}{\mathscr{I}}
\safemath{\colJ}{\mathscr{J}}
\safemath{\colK}{\mathscr{K}}
\safemath{\colL}{\mathscr{L}}
\safemath{\colM}{\mathscr{M}}
\safemath{\colN}{\mathscr{N}}
\safemath{\colO}{\mathscr{O}}
\safemath{\colP}{\mathscr{P}}
\safemath{\colQ}{\mathscr{Q}}
\safemath{\colR}{\mathscr{R}}
\safemath{\colS}{\mathscr{S}}
\safemath{\colT}{\mathscr{T}}
\safemath{\colU}{\mathscr{U}}
\safemath{\colV}{\mathscr{V}}
\safemath{\colW}{\mathscr{W}}
\safemath{\colX}{\mathscr{X}}
\safemath{\colY}{\mathscr{Y}}
\safemath{\colZ}{\mathscr{Z}}

\safemath{\opA}{\mathbb{A}}
\safemath{\opB}{\mathbb{B}}
\safemath{\opC}{\mathbb{C}}
\safemath{\opD}{\mathbb{D}}
\safemath{\opE}{\mathbb{E}}
\safemath{\opF}{\mathbb{F}}
\safemath{\opG}{\mathbb{G}}
\safemath{\opH}{\mathbb{H}}
\safemath{\opI}{\mathbb{I}}
\safemath{\opJ}{\mathbb{J}}
\safemath{\opK}{\mathbb{K}}
\safemath{\opL}{\mathbb{L}}
\safemath{\opM}{\mathbb{M}}
\safemath{\opN}{\mathbb{N}}
\safemath{\opO}{\mathbb{O}}
\safemath{\opP}{\mathbb{P}}
\safemath{\opQ}{\mathbb{Q}}
\safemath{\opR}{\mathbb{R}}
\safemath{\opS}{\mathbb{S}}
\safemath{\opT}{\mathbb{T}}
\safemath{\opU}{\mathbb{U}}
\safemath{\opV}{\mathbb{V}}
\safemath{\opW}{\mathbb{W}}
\safemath{\opX}{\mathbb{X}}
\safemath{\opY}{\mathbb{Y}}
\safemath{\opZ}{\mathbb{Z}}
\safemath{\opZero}{\mathbb{O}}
\safemath{\identityop}{\opI}

\safemath{\veca}{\bma}
\safemath{\vecb}{\bmb}
\safemath{\vecc}{\bmc}
\safemath{\vecd}{\bmd}
\safemath{\vece}{\bme}
\safemath{\vecf}{\bmf}
\safemath{\vecg}{\bmg}
\safemath{\vech}{\bmh}
\safemath{\veci}{\bmi}
\safemath{\vecj}{\bmj}
\safemath{\veck}{\bmk}
\safemath{\vecl}{\bml}
\safemath{\vecm}{\bmm}
\safemath{\vecn}{\bmn}
\safemath{\veco}{\bmo}
\safemath{\vecp}{\bmp}
\safemath{\vecq}{\bmq}
\safemath{\vecr}{\bmr}
\safemath{\vecs}{\bms}
\safemath{\vect}{\bmt}
\safemath{\vecu}{\bmu}
\safemath{\vecv}{\bmv}
\safemath{\vecw}{\bmw}
\safemath{\vecx}{\bmx}
\safemath{\vecy}{\bmy}
\safemath{\vecz}{\bmz}

\safemath{\veczero}{\bmzero}
\safemath{\vecone}{\bmone}
\safemath{\vecxi}{\bmxi}
\safemath{\veclambda}{\bmlambda}
\safemath{\vecmu}{\bmmu}
\safemath{\vectheta}{\bmtheta}
\safemath{\vecphi}{\bmphi}
\safemath{\vecdelta}{\bmdelta}

\safemath{\matA}{\bA}
\safemath{\matB}{\bB}
\safemath{\matC}{\bC}
\safemath{\matD}{\bD}
\safemath{\matE}{\bE}
\safemath{\matF}{\bF}
\safemath{\matG}{\bG}
\safemath{\matH}{\bH}
\safemath{\matI}{\bI}
\safemath{\matJ}{\bJ}
\safemath{\matK}{\bK}
\safemath{\matL}{\bL}
\safemath{\matM}{\bM}
\safemath{\matN}{\bN}
\safemath{\matO}{\bO}
\safemath{\matP}{\bP}
\safemath{\matQ}{\bQ}
\safemath{\matR}{\bR}
\safemath{\matS}{\bS}
\safemath{\matT}{\bT}
\safemath{\matU}{\bU}
\safemath{\matV}{\bV}
\safemath{\matW}{\bW}
\safemath{\matX}{\bX}
\safemath{\matY}{\bY}
\safemath{\matZ}{\bZ}
\safemath{\matzero}{\bmzero}

\safemath{\matDelta}{\bDelta}
\safemath{\matLambda}{\bLambda}
\safemath{\matPhi}{\bPhi}
\safemath{\matSigma}{\bSigma}
\safemath{\matOmega}{\bOmega}
\safemath{\matTheta}{\bTheta}

\safemath{\matidentity}{\matI}
\safemath{\matone}{\matO}

\safemath{\rnda}{A}
\safemath{\rndb}{B}
\safemath{\rndc}{C}
\safemath{\rndd}{D}
\safemath{\rnde}{E}
\safemath{\rndf}{F}
\safemath{\rndg}{G}
\safemath{\rndh}{H}
\safemath{\rndi}{I}
\safemath{\rndj}{J}
\safemath{\rndk}{K}
\safemath{\rndl}{L}
\safemath{\rndm}{M}
\safemath{\rndn}{N}
\safemath{\rndo}{O}
\safemath{\rndp}{P}
\safemath{\rndq}{Q}
\safemath{\rndr}{R}
\safemath{\rnds}{S}
\safemath{\rndt}{T}
\safemath{\rndu}{U}
\safemath{\rndv}{V}
\safemath{\rndw}{W}
\safemath{\rndx}{X}
\safemath{\rndy}{Y}
\safemath{\rndz}{Z}

\safemath{\rveca}{\bimA}
\safemath{\rvecb}{\bimB}
\safemath{\rvecc}{\bimC}
\safemath{\rvecd}{\bimD}
\safemath{\rvece}{\bimE}
\safemath{\rvecf}{\bimF}
\safemath{\rvecg}{\bimG}
\safemath{\rvech}{\bimH}
\safemath{\rveci}{\bimI}
\safemath{\rvecj}{\bimJ}
\safemath{\rveck}{\bimK}
\safemath{\rvecl}{\bimL}
\safemath{\rvecm}{\bimM}
\safemath{\rvecn}{\bimN}
\safemath{\rveco}{\bomO}
\safemath{\rvecp}{\bimP}
\safemath{\rvecq}{\bimQ}
\safemath{\rvecr}{\bimR}
\safemath{\rvecs}{\bimS}
\safemath{\rvect}{\bimT}
\safemath{\rvecu}{\bimU}
\safemath{\rvecv}{\bimV}
\safemath{\rvecw}{\bimW}
\safemath{\rvecx}{\bimX}
\safemath{\rvecy}{\bimY}
\safemath{\rvecz}{\bimZ}

\safemath{\rvecxi}{\bmxi}
\safemath{\rveclambda}{\bmlambda}
\safemath{\rvecmu}{\bmmu}
\safemath{\rvectheta}{\bmtheta}
\safemath{\rvecphi}{\bmphi}

\safemath{\rmatA}{\bimA}
\safemath{\rmatB}{\bimB}
\safemath{\rmatC}{\bimC}
\safemath{\rmatD}{\bimD}
\safemath{\rmatE}{\bimE}
\safemath{\rmatF}{\bimF}
\safemath{\rmatG}{\bimG}
\safemath{\rmatH}{\bimH}
\safemath{\rmatI}{\bimI}
\safemath{\rmatJ}{\bimJ}
\safemath{\rmatK}{\bimK}
\safemath{\rmatL}{\bimL}
\safemath{\rmatM}{\bimM}
\safemath{\rmatN}{\bimN}
\safemath{\rmatO}{\bimO}
\safemath{\rmatP}{\bimP}
\safemath{\rmatQ}{\bimQ}
\safemath{\rmatR}{\bimR}
\safemath{\rmatS}{\bimS}
\safemath{\rmatT}{\bimT}
\safemath{\rmatU}{\bimU}
\safemath{\rmatV}{\bimV}
\safemath{\rmatW}{\bimW}
\safemath{\rmatX}{\bimX}
\safemath{\rmatY}{\bimY}
\safemath{\rmatZ}{\bimZ}

\safemath{\rmatDelta}{\bimDelta}
\safemath{\rmatLambda}{\bimLambda}
\safemath{\rmatPhi}{\bimPhi}
\safemath{\rmatSigma}{\bimSigma}
\safemath{\rmatOmega}{\bimOmega}
\safemath{\rmatTheta}{\bimTheta}

\usepackage{amssymb}
\usepackage{amsfonts}
\usepackage{mathrsfs}
\usepackage{xspace}
\usepackage{bm}
\usepackage{fancyref}
\usepackage{textcomp}

\usepackage{multirow}
\usepackage{stmaryrd}

\newenvironment{textbmatrix}{	\setlength{\arraycolsep}{2.5pt}%
								\big[\begin{matrix}}{\end{matrix}\big]%
								\raisebox{0.08ex}{\vphantom{M}}}

\def\be{\begin{equation}}
\def\ee{\end{equation}}
\def\een{\nonumber \end{equation}}
\def\mat{\begin{bmatrix}}
\def\emat{\end{bmatrix}}
\def\btm{\begin{textbmatrix}}
\def\etm{\end{textbmatrix}}

\def\ba#1\ea{\begin{align}#1\end{align}}
\def\bas#1\eas{\begin{align*}#1\end{align*}}
\def\bs#1\es{\begin{split}#1\end{split}}
\def\bg#1\eg{\begin{gather}#1\end{gather}}
\def\bml#1\eml{\begin{multline}#1\end{multline}}
\def\bi#1\ei{\begin{itemize}#1\end{itemize}}

\newcommand{\lefto}{\mathopen{}\left}

\DeclareMathOperator*{\argmin}{arg\;min}		
\DeclareMathOperator*{\argmax}{arg\;max}		
\DeclareMathOperator{\Exop}{\opE}			

\newcommand{\Ex}[2]{\ensuremath{\Exop_{#1}\lefto[#2\right]}} 	

\newcommand{\tp}[1]{\ensuremath{#1^{\text{T}}}} 		
\newcommand{\herm}[1]{\ensuremath{#1^{\text{H}}}} 	
\newcommand{\inv}[1]{\ensuremath{#1^{-1}}} 	
\newcommand{\pinv}[1]{\ensuremath{#1^{\dagger}}} 	

\safemath{\dirac}{\delta}					
\safemath{\krond}{\dirac}					

\safemath{\upto}{\uparrow}
\safemath{\downto}{\downarrow}
\safemath{\iu}{j}							
\safemath{\ev}{\lambda}						
\safemath{\hilseqspace}{l^{2}}				
\newcommand{\banachfunspace}[1]{\setL^{#1}}	
\safemath{\hilfunspace}{\banachfunspace{2}}	

\safemath{\SNR}{\textit{SNR}} 				
\safemath{\PAR}{\textit{PAR}} 				
\safemath{\No}{N_0}							
\safemath{\Es}{E_s}							
\safemath{\Eb}{E_b}							
\safemath{\EbNo}{\frac{\Eb}{\No}}
\safemath{\EsNo}{\frac{\Es}{\No}}

\DeclareMathOperator{\CHop}{\ensuremath{\opH}} 
\safemath{\tvir}{\rndh_{\CHop}}				
\safemath{\tvtf}{\rndl_{\CHop}}				
\safemath{\spf}{\rnds_{\CHop}}				
\safemath{\bff}{H_{\CHop}}					

\safemath{\ircf}{r_{h}}						
\safemath{\tftvcf}{r_{s}}					
\safemath{\tfcf}{r_{l}}						
\safemath{\bfcf}{r_{H}}						

\safemath{\tcorr}{c_h}						
\safemath{\scf}{c_{s}}						
\safemath{\tfcorr}{c_{l}}					
\safemath{\fcorr}{c_{H}}						

\safemath{\mi}{I}							
\safemath{\capacity}{C}						

\safemath{\normal}{\mathcal{N}}			
\safemath{\jpg}{\mathcal{CN}}			
\safemath{\mchain}{\leftrightarrow}		

\safemath{\dB}{\,\mathrm{dB}}
\safemath{\dBm}{\,\mathrm{dBm}}
\safemath{\Hz}{\,\mathrm{Hz}}
\safemath{\kHz}{\,\mathrm{kHz}}
\safemath{\MHz}{\,\mathrm{MHz}}
\safemath{\GHz}{\,\mathrm{GHz}}
\safemath{\s}{\,\mathrm{s}}
\safemath{\ms}{\,\mathrm{ms}}
\safemath{\mus}{\,\mathrm{\text{\textmu}s}}
\safemath{\ns}{\,\mathrm{ns}}
\safemath{\ps}{\,\mathrm{ps}}
\safemath{\meter}{\,\mathrm{m}}
\safemath{\mm}{\,\mathrm{mm}}
\safemath{\cm}{\,\mathrm{cm}}
\safemath{\m}{\,\mathrm{m}}
\safemath{\W}{\,\mathrm{W}}
\safemath{\mW}{\, \mathrm{mW}}
\safemath{\J}{\,\mathrm{J}}
\safemath{\K}{\,\mathrm{K}}
\safemath{\bit}{\,\mathrm{bit}}
\safemath{\nat}{\,\mathrm{nat}}

\safemath{\define}{\triangleq}			

\safemath{\equivalent}{\sim}
\safemath{\distas}{\sim}					
\safemath{\sdiff}{\Delta}				

\safemath{\reals}{\mathbb{R}}
\safemath{\positivereals}{\reals_{+}}
\safemath{\integers}{\mathbb{Z}}
\safemath{\posint}{\integers_{+}}
\safemath{\naturals}{\mathbb{N}}
\safemath{\posnaturals}{\naturals_{+}}
\safemath{\complexset}{\mathbb{C}}
\safemath{\rationals}{\mathbb{Q}}

\newcommand*{\fancyrefapplabelprefix}{app}		
\newcommand*{\fancyrefthmlabelprefix}{thm}		
\newcommand*{\fancyreflemlabelprefix}{lem}		
\newcommand*{\fancyrefcorlabelprefix}{cor}		
\newcommand*{\fancyrefdeflabelprefix}{def}		
\newcommand*{\fancyrefproplabelprefix}{prop}		
\newcommand*{\fancyrefexmpllabelprefix}{exmpl}
\newcommand*{\fancyrefalglabelprefix}{alg}		
\newcommand*{\fancyreftbllabelprefix}{tbl}		

\frefformat{vario}{\fancyrefseclabelprefix}{Section~#1}
\frefformat{vario}{\fancyrefthmlabelprefix}{Theorem~#1}
\frefformat{vario}{\fancyreftbllabelprefix}{Table~#1}
\frefformat{vario}{\fancyreflemlabelprefix}{Lemma~#1}
\frefformat{vario}{\fancyrefcorlabelprefix}{Corollary~#1}
\frefformat{vario}{\fancyrefdeflabelprefix}{Definition~#1}
\frefformat{vario}{\fancyreffiglabelprefix}{Fig.~#1}
\frefformat{vario}{\fancyrefapplabelprefix}{Appendix~#1}
\frefformat{vario}{\fancyrefeqlabelprefix}{(#1)}
\frefformat{vario}{\fancyrefproplabelprefix}{Proposition~#1}
\frefformat{vario}{\fancyrefexmpllabelprefix}{Example~#1}
\frefformat{vario}{\fancyrefalglabelprefix}{Algorithm~#1}

\safemath{\dictab}{[\,\dicta\,\,\dictb\,]}

\safemath{\ysig}{\bmy}
\safemath{\ysighat}{\hat{\ysig}}
\safemath{\ysigdim}{M}
\safemath{\xsig}{\bmx}
\safemath{\xsigdim}{N}
\safemath{\nx}{n_x}
\safemath{\zsig}{\bmz}
\safemath{\zsigdim}{\ysigdim}
\safemath{\rsig}{\bmr}
\safemath{\Adict}{\bA}
\safemath{\Adicttilde}{\widetilde{\Adict}}
\safemath{\Adictdim}{\outputdim\times\xsigdim}
\safemath{\avec}{\bma}
\safemath{\avectilde}{\tilde{\avec}}
\safemath{\Bdict}{\bB}
\safemath{\Bdicttilde}{\widetilde{\Bdict}}
\safemath{\Cdict}{\bC}
\safemath{\cvec}{\bmc}
\safemath{\Ddict}{\bD}
\safemath{\Ddictdim}{\ysigdim\times\xsigdim}
\safemath{\dvec}{\bmd}
\safemath{\Ddicttilde}{\widetilde{\bD}}
\safemath{\Bonb}{\bB}
\safemath{\bvec}{\bmb}
\safemath{\Bonbdim}{\ysigdim\times\ysigdim}
\safemath{\noise}{\bmn}
\safemath{\noisedim}{\ysigim}
\safemath{\err}{\bme}
\safemath{\errdim}{\ysigdim}
\safemath{\errset}{\setE}
\safemath{\nerr}{n_e}
\safemath{\delop}{\bP_\errset}
\safemath{\delopc}{\bP_{{\errset}^c}}

\safemath{\cplxi}{\imath}
\safemath{\cplxj}{\jmath}

\safemath{\dict}{\matD}
\safemath{\inputdim}{N}		
\safemath{\outputdim}{M}		
\safemath{\sparsity}{S}	
\safemath{\inputdimA}{{N_a}}	
\safemath{\inputdimB}{{N_b}}	
\safemath{\elemA}{{n_a}}	
\safemath{\elemB}{{n_b}}	
\safemath{\resA}{\matR_a}	
\safemath{\resB}{\matR_b}	
\safemath{\subD}{\matS} 
\safemath{\subA}{\matS_a} 
\safemath{\subB}{\matS_b} 
\safemath{\dicta}{\matA} 	
\safemath{\dictb}{\matB} 	
\safemath{\hollowS}{H}
\safemath{\hollowA}{H_a}
\safemath{\hollowB}{H_b}
\safemath{\cross}{Z}
\safemath{\coh}{\mu_d}			
\safemath{\coha}{\mu_a}			
\safemath{\cohb}{\mu_b}			
\safemath{\mubs}{\nu}	
\safemath{\cohm}{\mu_m} 
\safemath{\dictset}{\setD}	
\safemath{\dictsetp}{\dictset(\coh,\coha,\cohb)}	
\safemath{\dictsetgen}{\dictset_\text{gen}}
\safemath{\dictsetgenp}{\dictsetgen(\coh)}
\safemath{\dictsetonb}{\dictset_\text{onb}}
\safemath{\dictsetonbp}{\dictsetonb(\coh)}

\safemath{\leftside}{U}
\safemath{\rightsideA}{R_a}
\safemath{\rightsideB}{R_b}

\safemath{\indexS}{\setI_S} 

\safemath{\na}{n_a}			
\safemath{\nb}{n_b}			
\safemath{\coeffa}{p_i}	
\safemath{\coeffb}{q_j}	
\safemath{\seta}{\setP}		
\safemath{\setb}{\setQ}     
\safemath{\setw}{\setW}	
\safemath{\setz}{\setZ}	
\safemath{\cola}{\veca}		
\safemath{\colb}{\vecb}		
\safemath{\cold}{\vecd}		
\safemath{\inputvec}{\vecx} 	
\safemath{\error}{\vece}	
\safemath{\noiseout}{\vecz} 	
\safemath{\inputvecel}{x}
\safemath{\inputveca}{\vecx_a}
\safemath{\inputvecb}{\vecx_b}
\safemath{\outputvec}{\vecy}	
\safemath{\lambdamin}{\lambda_{\mathrm{min}}}

\safemath{\elltwo}{\ell_2}
\safemath{\ellone}{\ell_1}
\safemath{\ellzero}{\ell_0}
\safemath{\ellinf}{\ell_\infty}
\safemath{\ellinftilde}{\ell_{\widetilde\infty}}
\safemath{\licard}{Z(\coh,\coha,\cohb)}
\safemath{\xsol}{\hat{x}}
\safemath{\xbord}{x_b}		
\safemath{\xstat}{x_s}		
\safemath{\xstatLone}{\tilde{x}_s}
\safemath{\order}{\mathcal{O}} 
\safemath{\scales}{\Theta} 
\safemath{\ones}{\mathbf{1}} 
\safemath{\zeroes}{\mathbf{0}} 
\safemath{\thlone}{\kappa(\coh,\cohb)} 
\safemath{\constoneA}{\delta} 
\safemath{\constoneB}{\epsilon} 
\safemath{\nlarge}{L}				   
\safemath{\sumlarge}{S_\nlarge}
\safemath{\maxlarger}{P_\nlarge}	   
\safemath{\Pzero}{\textrm{P0}}	
\safemath{\Pone}{\textrm{P1}}
\safemath{\vecfir}{\vecw}			 
\safemath{\vecsec}{\vecz}
\safemath{\elvecfir}{w}              
\safemath{\elvecsec}{z}				 
\safemath{\nlargefir}{n}
\safemath{\normout}{\gamma}
\safemath{\auxfun}{h}
\safemath{\supp}{\textrm{supp}}

\safemath{\indexa}{\ell}
\safemath{\indexb}{r}
\safemath{\indexc}{i}
\safemath{\indexd}{j}

\safemath{\project}{P}

\usepackage{framed}

\IEEEoverridecommandlockouts
\allowdisplaybreaks 

\safemath{\Hj}{\bmj}
\safemath{\bsj}{\bmw}
\safemath{\sj}{w}
\safemath{\Ej}{E_w}
\safemath{\proxg}{\text{prox}_g}
\safemath{\rE}{\rho_{\textsf{\tiny{E}}}}
\safemath{\rP}{\rho_{\textsf{\tiny{P}}}}

\begin{document}
\bstctlcite{IEEEexample:BSTcontrol} 

\title{Mitigating Smart Jammers in MU-MIMO via\\Joint Channel Estimation and Data Detection}

\author{\IEEEauthorblockN{Gian Marti and Christoph Studer}
\IEEEauthorblockA{\em Department of Information Technology
and Electrical Engineering, ETH Zurich, Switzerland\\
e-mail: gimarti@ethz.ch and  studer@ethz.ch\vspace{-7mm}
}
\thanks{The work of CS was supported in part by ComSenTer, one of six centers in JUMP, a SRC program sponsored by DARPA, 
and in part by the U.S.\ National Science Foundation (NSF) under grants CNS-1717559 and ECCS-1824379. The work of GM and CS was supported in part by an ETH Research~Grant.}
\thanks{The authors thank O. Casta\~neda and S. Taner for comments and suggestions.}
}

\maketitle

\begin{abstract}
Wireless systems must be resilient to jamming attacks. 
Existing mitigation methods require knowledge of the jammer's  transmit characteristics. 
However, this knowledge may be difficult to acquire, especially for smart jammers that attack only specific instants during transmission in order to evade mitigation.
We propose a novel method that mitigates attacks by smart jammers on massive multi-user multiple-input multiple-output (MU-MIMO) basestations (BSs).
Our approach builds on recent progress in joint channel estimation and data detection (JED) and 
exploits the fact that a jammer cannot change its subspace within a coherence interval. 
Our method, called MAED (short for MitigAtion, Estimation, and~Detection), uses a novel problem 
formulation that combines jammer estimation and mitigation, channel estimation, and data detection, instead of separating these tasks.
We solve the problem approximately with an efficient iterative algorithm. 
Our results show that MAED effectively mitigates a wide range
of smart jamming attacks without having any \textit{a priori} knowledge about the attack type.
\end{abstract}

\section{Introduction}
Jamming attacks pose a serious threat to the continuous operability of wireless communication systems \cite{economist2021satellite}.
Effective methods to mitigate such attacks are necessary
as wireless systems become increasingly critical to modern infrastructure~\cite{popovski2014ultra}.
In the uplink of massive multi-user multiple-input multiple-output (MU-MIMO) systems, effective jammer mitigation is rendered possible by the asymmetry in the number of antennas between the basestation (BS), which has many antennas, and a mobile jamming device, which has one or few antennas.
One possibility, for instance, is to project the receive signals on the subspace 
orthogonal to the jammer's channel~\cite{marti2021snips,yan2016jamming}. 
But such methods require accurate knowledge of the jammer's channel.
If a jammer transmits permanently and with a static signature (often called barrage jamming), the~BS~can estimate the required quantities, for instance during a dedicated period in which the user equipments (UEs) do not transmit~\cite{marti2021snips} or in which they transmit predefined symbols~\cite{yan2016jamming}.
Instead of barrage jamming, however, a smart jammer might jam the system only at specific time instants.
Such attacks may prevent the BS from estimating a jammer's channel with simple~methods.
\nopagebreak
\subsection{State of the Art}

\begin{figure}[tp]
\vspace{-1mm}
\centering
~
\subfigure[Existing jammer-mitigation methods separate jammer estimation (JEST) and channel estimation (CHEST) from the jammer-resilient
data detection (DET). Such methods are~ineffective~against~jammers that attack only the data~phase and do not transmit in the training phase.]{
\includegraphics[width=0.956\columnwidth]{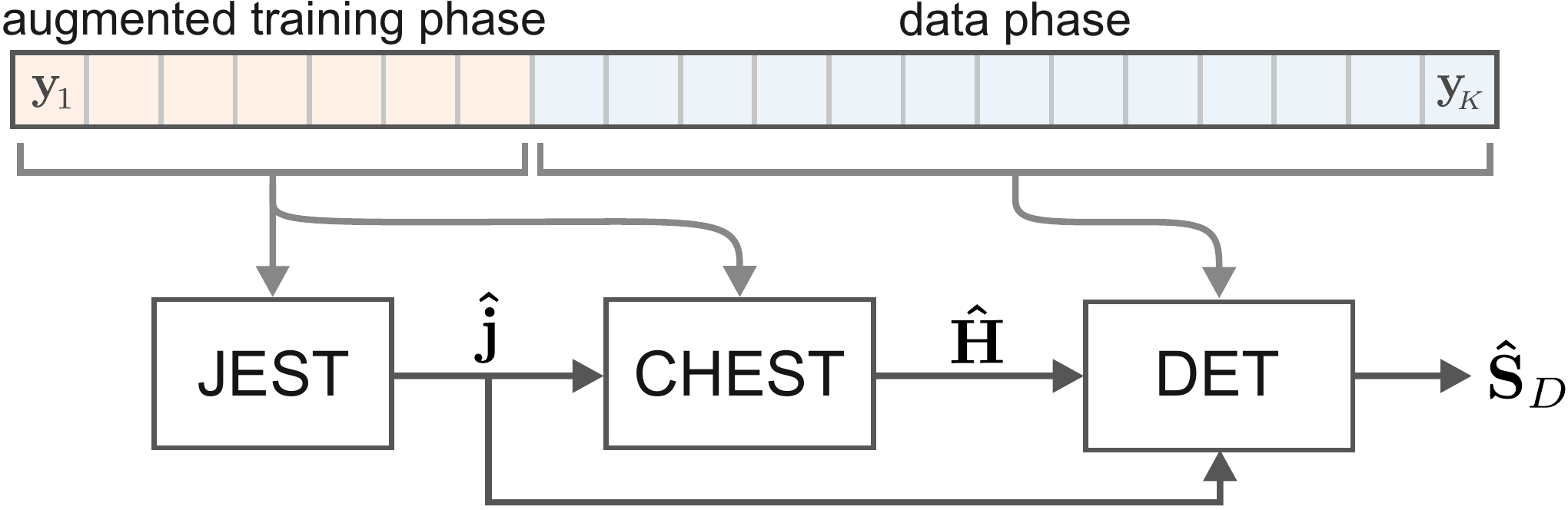}
\label{fig:traditional}
}
\newline
\subfigure[MAED unifies jammer estimation and mitigation, channel estimation, and data detection 
to mitigate jamming attacks regardless of their activity~pattern.]{
\includegraphics[width=0.956\columnwidth]{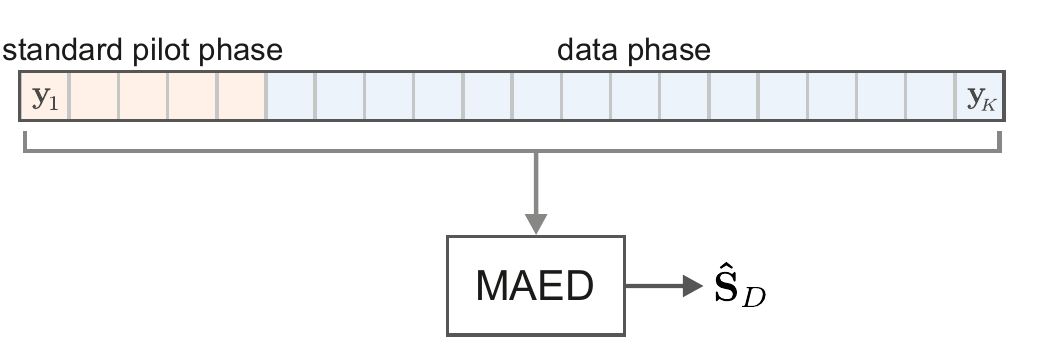}
\label{fig:maed}
}
\caption{
The approach to jammer mitigation taken by existing methods (a) compared to the approach taken by MAED (b).
In the figure, $\{\bmy_1,\dots,\bmy_K\}$ are the receive signals, 
and $\hat\Hj, \hat\bH$ and $\hat\bS_D$ are the estimates of the jammer channel, the UE channel matrix, and the UE transmit symbols, respectively.
}
\label{fig:maed_vs_trad}
\end{figure}

Multi-antenna wireless systems have the unique potential to effectively mitigate jamming attacks,
and a variety of multi-antenna methods have been proposed for the mitigation of jamming attacks in MIMO systems
\cite{marti2021snips, shen14a, yan2016jamming,zeng2017enabling, vinogradova16a, do18a, akhlaghpasand20a, akhlaghpasand20b, marti2021hybrid}.
Common to all of them~is the assumption---in one way or other---that information about the jammer's transmit characteristics
(e.g., the jammer's channel, or the covariance matrix between the UE transmit signals and the jammed receive signals)
can be estimated based on some specific subset of the receive samples.\footnote{The method of \cite{vinogradova16a} is to some extent an exception as it estimates the~UEs' subspace and projects the receive signals thereon. However, this method~dist-inguishes the UEs' from the jammer's subspace based on the receive power, thereby presuming that the UEs and the jammer transmit with different power.}
\fref{fig:traditional}~illustrates the approach taken by these methods: 
The data phase is preceded by an augmented 
training phase in which the jammer's transmit characteristics are estimated in addition to the channel matrix. 
This augmented training phase can either complement a traditional pilot phase with a period during which the UEs do not transmit in order to 
enable jammer estimation (e.g., \cite{marti2021snips, shen14a}), or it can consist of an extended pilot phase so that there exist pilot sequences which are unused by
the UEs and on whose span the receive signals can be projected to estimate the jammer's subspace (e.g., \cite{do18a, akhlaghpasand20a, akhlaghpasand20b}). 
The estimated jammer characteristics are then used to perform jammer-mitigating data detection. 
Such an approach succeeds in the case of barrage jammers, but is unreliable for estimating the transmit characteristics of smart jammers: 
A smart jammer can evade estimation and thus circumvent mitigation by not~transmitting in those samples, for instance because it is aware of the defense mechanism, 
or simply because it jams in brief bursts only. 
For this reason, our proposed method MAED  unifies jammer estimation, channel estimation and data detection, see \fref{fig:maed}.

Many studies have already shown how smart jammers can disrupt wireless communication systems by
targeting only specific parts
of the wireless transmission \cite{miller2010subverting, miller2011vulnerabilities, clancy2011efficient, sodagari2012efficient, 
lichtman2013vulnerability, lichtman20185g, lichtman2016lte, girke2019towards,lapan2012jamming} 
instead of using barrage jamming.
Jammers that jam only the pilot phase have received considerable attention 
\cite{miller2010subverting,miller2011vulnerabilities,clancy2011efficient,sodagari2012efficient,lichtman2013vulnerability}, 
as such attacks can be more energy-efficient than barrage jamming in disrupting communication systems that do not 
defend themselves against jammers~\cite{clancy2011efficient,sodagari2012efficient,lichtman2013vulnerability}.
However, if a jammer is active during the pilot phase, then a BS that \emph{does} defend itself against jamming attacks
can estimate the jammer's channel by exploiting knowledge of the UE transmit symbols, for instance with the aid of unused pilot sequences \cite{do18a, akhlaghpasand20a, akhlaghpasand20b}.
To disable such jammer-mitigating communication systems, a smart jammer might therefore refrain from jamming the pilot phase and only target 
the data phase, even if such jamming attacks have not received much attention so far 
\cite{lichtman2016lte, girke2019towards}.
Other threat models that have been analyzed include jammers that attack specific control 
\mbox{channels \cite{lichtman2013vulnerability, lichtman2016lte, lichtman20185g} or
the time synchronization phase \cite{lapan2012jamming}.}

\subsection{Contributions}
We propose MAED (short for MitigAtion, Estimation, and Detection), 
a novel method for jammer mitigation that does not depend on the jammer being active during 
\textit{any} specific period.
Instead, our approach leverages the fact that a jammer cannot change its subspace instantaneously. 
To this end, MAED utilizes a problem formulation which unifies jammer estimation and cancellation, 
channel estimation, and data detection, instead of dealing with these tasks independently, see \fref{fig:maed}.
In doing this, the method builds upon techniques for joint channel estimation and data detection (JED)
\cite{vikalo2006efficient, xu2008exact, kofidis2017joint, castaneda2018vlsi, yilmaz2019channel, he2020model, song2021soft}. 
The proposed problem formulation is approximately solved using an efficient iterative algorithm. 
Extensive simulation results demonstrate that MAED is able to effectively mitigate a wide variety of na\"ive and smart  jamming attacks,
without requiring any \textit{a~priori} knowledge about the attack type.

\subsection{Notation}
Matrices and column vectors are represented by boldface uppercase and lowercase letters, respectively.
For a matrix $\bA$, the transpose is $\tp{\bA}$, the conjugate transpose is $\herm{\bA}$, the entry in the $\ell$th row and $k$th column is $[\bA]_{\ell,k}$, 
the submatrix consisting of the columns from $n$ through $m$ is $\bA_{[n:m]}$,
and the Frobenius norm is $\| \bA \|_F$.
The $N\!\times\!N$ identity matrix is $\bI_N$.
For a vector~$\bma$, the $\ell_2$-norm is $\|\bma\|_2$, the real part is $\Re\{\bma\}$, and the imaginary part is $\Im\{\bma\}$.
The expectation with respect to the random vector~$\bmx$ is denoted by \Ex{\bmx}{\cdot}.
We define $i^2=-1$.
The complex $n$-hypersphere of radius $r$ is denoted by $\mathbb{S}_r^n$,
and the span of a vector $\bma$ is denoted by $\textit{span}(\bma)$.

\section{System Setup}\label{sec:setup}
We consider the uplink of a massive MU-MIMO system in which $U$ single-antenna UEs transmit data to 
a $B$ antenna BS in the presence of a single-antenna jammer. 
The channels are assumed to be frequency flat and block-fading with coherence time $K=T+D$.
The first $T$ time slots are used to 
transmit pilot symbols; the remaining $D$ time slots are used to transmit payload data.
The UE transmit matrix is $\bS = [\bS_T,\bS_D]$, where $\bS_T\in \opC^{U\times T}$
and $\bS_D\in\setS^{U\times D}$ contain the pilots and the transmit symbols, respectively,
and $\setS$ is the transmit constellation, which is taken to be QPSK with symbol energy $\Es$
throughout this 
paper.\footnote{Our method can be extended to higher-order modulation schemes.}
We assume that the jammer does not prevent the UEs and the BS from establishing time synchronization,
which allows us to use the discrete-time input-output relation
\begin{align}
	\bY = \bH\bS + \Hj\tp{\bsj} + \bN. \label{eq:io}
\end{align}
Here, $\bY\in\opC^{B\times K}$ is the BS-side receive matrix that contains the \mbox{$B$-dimensional} receive vectors over all $K$ time slots, 
\mbox{$\bH\in\opC^{B\times U}$} models the MIMO uplink channel,
$\Hj\in\opC^B$ models the channel between the jammer and the BS, $\bsj\in\opC^K$  contains the jammer's transmit symbols over all $K$ time slots, 
and $\bN\in\opC^{B\times K}$ models thermal noise and contains independently and identically distributed (i.i.d.) circularly-symmetric complex Gaussian entries with variance $N_0$.
In what follows, we usually assume that the jammer's transmit symbols $\bsj$ are independent of the
UE transmit symbols $\bS$. However, we will also consider a scenario in which we drop this assumption (see \fref{sec:smart}). 
We make no other assumptions about the distribution of $\bsj$. 
In particular, we do not assume that its entries are distributed~i.i.d.

\section{Joint Jammer Estimation and Mitigation, Channel Estimation, and Data Detection}

Our goal is threefold: (i) mitigating the jammer's interference by locating its one-dimensional subspace 
$\textit{span}(\Hj)$
and projecting the receive matrix $\bY$ onto the orthogonal complement of~that subspace, 
(ii) estimating the channel matrix $\bH$, and (iii) recovering the UE data in $\bS_D$.
To this end, we develop a novel problem formulation that combines all three of these aspects. 
We then propose an iterative algorithm to approximately solve this optimization problem. 

\subsection{The  MAED Optimization Problem}

In the absence of a jammer, the maximum-likelihood problem for joint channel estimation 
and data detection is \cite{vikalo2006efficient}
\begin{align}
	 \big\{\hat\bH, \hat\bS_D\big\}
	&= \argmin_{\substack{\hspace{1.3mm}\tilde\bH\in\opC^{B\times U}\\ \tilde\bS_D\in\setS^{U\times D}}}\!
	\big\|\bY - \tilde\bH \tilde\bS \big\|^2_F, \label{eq:ml_jed}
\end{align}
where, for brevity, we define $\tilde\bS \triangleq [\bS_T,\tilde\bS_D]$ and leave the dependence on
$\tilde\bS_D$ implicit. This objective already integrates the goals of estimating the channel matrix
and detecting the transmit symbols.
The jammer, however, will lead to a residual 
\begin{align}
	\|\bY - \bH\bS\|^2_F &= \|\Hj\tp{\bsj} + \bN\|^2_F \approx \|\Hj\tp{\bsj}\|^2_F
\end{align}
when plugging the true channel and data matrices into \fref{eq:ml_jed}, and assuming that the contribution of the
noise $\bN$ is negligible.

Consider now the projection onto the orthogonal subspace (POS) for jammer mitigation \cite{marti2021snips}: 
POS nulls a jammer by orthogonally projecting the receive signals onto the orthogonal complement of $\textit{span}(\Hj)$ using the 
matrix $\bP(\Hj) \triangleq \bI_B - \Hj\herm{\Hj}/\|\Hj\|^2$:
\begin{align}
	\bP(\Hj)\bY 
	&= \bP(\Hj)\,\bH\bS + \bP(\Hj)\,\Hj\tp{\bsj} + \bP_{\Hj}\,\bN \label{eq:pos} \\
	&= \bP(\Hj)\,\bH\bS + \bP(\Hj)\,\bN.
\end{align}
One can then define \mbox{$\bY_{\bP(\Hj)}\triangleq\bP(\Hj)\bY,~\bH_{\bP(\Hj)}\triangleq\bP(\Hj)\bH$,}
and perform channel estimation and data detection using the resulting jammer-free system.
The difficulty is, of course, that the projection matrix $\bP(\Hj)$ depends on the (unknown) direction~$\Hj/\|\Hj\|$ of the jammer's channel.

Now, consider what happens when we take the matrix\footnote{The dependence
of $\tilde\bP$ on $\tilde\bmp$ is left implicit here and throughout the paper.} 
$\tilde\bP\triangleq\bI-\tilde\bmp\herm{\tilde\bmp},~\tilde\bmp\in \mathbb{S}_1^B$ which orthogonally projects a signal 
onto the orthogonal complement of some arbitrary one-dimen-sional subspace $\textit{span}(\tilde\bmp)$,
and then apply that projection to the objective of \eqref{eq:ml_jed} as follows: 
\begin{align}
	\|\tilde\bP(\bY - \tilde\bH\tilde\bS)\|^2_F. \label{eq:ml_p_jed}
\end{align}
If we now plug the true channel and data matrices into \fref{eq:ml_p_jed} and assume that the noise $\bN$ is negligible, then we obtain
\begin{align}
	\|\tilde\bP(\bY - \bH\bS)\|^2_F 
	&= \|\tilde\bP\Hj\tp{\bsj} + \tilde\bP\bN\|^2_F \\
	&\approx  \|\tilde\bP\Hj\tp{\bsj}\|^2_F \\ 
	&\geq 0,
\end{align}
with equality if and only if $\tilde\bmp$ is collinear with $\Hj$. 
In other words, the unit vector $\tilde\bmp$ which---in combination with the true channel and data matrices---minimizes \eqref{eq:ml_p_jed}
is collinear with the jammer's channel, and hence $\tilde\bP$ is the POS matrix.

So, assuming the noise $\bN$ is negligible, then $\{\tilde\bmp,\tilde\bH,\tilde\bS\}$ minimizes \eqref{eq:ml_p_jed} if
(i) $\tilde\bP$ is the orthogonal projection onto~the orthogonal complement of $\textit{span}(\Hj)$,
(ii) $\tilde\bH$ is the true channel matrix,
and (iii) $\tilde\bS$ contains the true data matrix.
These~are, of course, exactly the goals that we wanted to attain.
Following this insight, we now formulate the MAED joint jammer estimation and mitigation, channel estimation, \mbox{and data detection~problem:}
\begin{align}
	 \big\{\hat\bmp, \hat\bH_\bP, \hat\bS_D\big\}
	&= \argmin_{\substack{\tilde\bmp\in \mathbb{S}_1^B\hspace{1.4mm}\\ \hspace{1.3mm}\tilde\bH_\bP\in\opC^{B\times U}\\ \tilde\bS_D\in\setS^{U\times D}}}\!
	\big\|\tilde\bP\bY - \tilde\bH_\bP \tilde\bS \big\|^2_F.\!
	\label{eq:obj1}
\end{align}
Note that, compared to \eqref{eq:ml_p_jed}, we have absorbed the projection matrix $\tilde\bP$ directly into 
the unknown channel matrix $\tilde\bH_\bP$. 
Otherwise the columns of $\tilde\bH_\bP$ would be ill-defined with respect to the length of their components in the direction of $\tilde\bmp\approx\Hj$, meaning that 
one could not distinguish between channel estimates $\tilde\bH + \alpha\Hj\tp{\tilde{\bsj}}$
with different $\alpha,\tilde{\bsj}$.

\subsection{Solving the MAED Optimization Problem}
The objective \eqref{eq:obj1} is quadratic in $\tilde\bH_\bP$, 
so we can derive the optimal value of $\tilde\bH_\bP$ as a function of $\tilde\bP$ and $\tilde\bS$, as
\begin{align}
	\hat\bH_\bP = \tilde\bP\bY\pinv{\tilde\bS}, 
\end{align}
where $\pinv{\tilde\bS}=\herm{\tilde\bS}\inv{(\tilde\bS\herm{\tilde\bS})}$ is the Moore-Penrose
pseudo-inverse of $\tilde\bS$. Substituting $\hat\bH_\bP$ back into \eqref{eq:obj1} yields 
\begin{align}
	\big\{\hat\bmp, \hat\bS_D\big\} = 
	\argmin_{\substack{\tilde\bmp\in \mathbb{S}_1^B\hspace{1.4mm}\\ \tilde\bS_D\in\setS^{D\times U}}}
	\big\|\tilde\bP\bY(\bI_K - \pinv{\tilde\bS}\tilde\bS)\big\|^2_F. \label{eq:obj3}
\end{align}
Solving \eqref{eq:obj3} is difficult due to its combinatorial nature, so we resort to solving it approximately. 
First, we relax the constraint set $\setS$ to its convex hull $\setC\triangleq\textit{conv}(\setS)$ as in \cite{castaneda2018vlsi}.
We then solve this~relaxed problem formulation approximately by alternately performing 
a forward-backward splitting step in $\tilde\bS$ and a minimization step in $\tilde\bP$.

\subsection{Forward-Backward Splitting Step in $\tilde\bS$} \label{sec:fbs}
Forward-backward splitting (FBS) \cite{goldstein16a}, also called proximal gradient descent,
is an iterative method that solves convex optimization problems of the form
\begin{align}
	\argmin_{\tilde\bms}\, f(\tilde\bms) + g(\tilde\bms), \label{eq:fbs1}
\end{align}
where $f$ is convex and differentiable, and $g$ is convex but not necessarily
differentiable, smooth, or bounded. Starting from an initialization vector $\tilde\bms^{(0)}$, 
FBS solves the problem in~\eqref{eq:fbs1} iteratively by computing 
\begin{align}
	\tilde\bms^{(t+1)} = \proxg\big(\tilde\bms^{(t)} - \tau^{(t)}\nabla f(\tilde\bms^{(t)}); \tau^{(t)}\big).
\end{align}
Here, $\nabla f(\tilde\bms)$ is the gradient of $f(\tilde\bms)$, $\tau^{(t)}$ is the stepsize at iteration $t$, 
and $\proxg$ is the proximal operator of $g$ \cite{parikh13a}.
For a suitably chosen sequence of stepsizes $\{\tau^{(t)}\}$, FBS solves convex optimization problems exactly
(provided that the number of iterations is sufficiently large). 
FBS can also be utilized to approximately and efficiently solve non-convex
problems, even though there are typically no guarantees for optimality or even convergence~\cite{goldstein16a}.

For the optimization problem in \fref{eq:obj3}, we define $f$ and $g$ as 
\begin{align}
	f(\tilde\bS) &= \big\|\tilde\bP\bY(\bI_K - \pinv{\tilde\bS}\tilde\bS)\big\|^2_F,
\end{align}
and
\begin{align}
	g(\tilde\bS) &= \begin{cases}
		0 &\text{if }\,\tilde\bS_{[1:T]}=\bS_T \text{ and } \tilde\bS_{[T+1:K]}\in\setC^{U\times D}
		\!\!\!\\
		\infty &\text{else}.
	\end{cases}
\end{align}
The gradient of $f$ in $\tilde\bS$ is given by 
\begin{align}
	\nabla f(\tilde\bS) = \herm{(\pinv{\tilde\bS})}\herm\bY\tilde\bP\bY(\bI_K - \pinv{\tilde\bS}\tilde\bS)
\end{align}
and the proximal operator for $g$ is simply the orthogonal projection onto the constraint set, 
which is 
\begin{align}
	[\proxg(\tilde\bS)]_{u,k} = \begin{cases}
		[\bS_T]_{u,k} &\text{ if } k\in[1:T] \\
		\text{proj}_\setC([\tilde\bS_{u,k}]) &\text{ else,}
	\end{cases} \label{eq:proxg} 
\end{align}
where (for QPSK) $\text{proj}_\setC$ acts entry-wise on $[\tilde\bS]_{u,k}$ as 
\begin{align}
	\text{proj}_\setC(x) =\, & \min\{\max\{\Re(x),-\sqrt{\sfrac{\Es}{2}}\},\sqrt{\sfrac{\Es}{2}}\} \nonumber\\
	&+ i\min\{\max\{\Im(x),-\sqrt{\sfrac{\Es}{2}}\},\sqrt{\sfrac{\Es}{2}}\}.
\end{align}
For the selection of the stepsizes $\{\tau^{(t)}\}$, we use the Barzilai-Borwein method 
\cite{barzilai1988two} as detailed in \cite{goldstein16a,zhou2006gradient}.

\subsection{Minimization Step in $\tilde\bP$}
After the FBS step in $\tilde\bS$, we minimize \eqref{eq:obj3}
with respect to the vector~$\tilde\bmp$. Defining 
$\tilde\bE\triangleq \bY(\bI_K - \pinv{\tilde{\bS}}\tilde{\bS})$
and performing standard algebraic manipulations yields
\begin{align}
	\hat\bmp &= \argmin_{\tilde\bmp\in \mathbb{S}_1^B} \big\|\tilde\bP \tilde\bE \big\|^2_F	\\
	&= \argmax_{\tilde\bmp\in \mathbb{S}_1^B} \, \herm{\tilde\bmp} \tilde\bE \herm{\tilde\bE} \tilde\bmp. \label{eq:rayleigh}
\end{align}
This implies that the minimizer $\hat\bmp$ is the unit vector which maximizes the 
Rayleigh quotient of $\tilde\bE \herm{\tilde\bE}$. The~solution to~this problem is the eigenvector 
$\bmv_1(\tilde\bE \herm{\tilde\bE})$ belonging~to~the largest eigenvalue of $\tilde\bE \herm{\tilde\bE}$, normalized to unit length~\cite[Thm.\,4.2.2]{horn2013matrix},
\begin{align}
	\hat\bmp=\frac{\bmv_1(\tilde\bE \herm{\tilde\bE})}{\big\|\bmv_1(\tilde\bE \herm{\tilde\bE})\big\|_2}.
\end{align}
Calculating this eigenvector for every iteration is computationally expensive, 
so we only do it for the very first iteration. 
In~all subsequent iterations, we then approximate its value with a single power method step \cite[Sec.\,8.2.1]{GV96}, i.e., 
we estimate 
\begin{align}
	\hat\bmp^{(t+1)} = \frac{\tilde\bE^{(t+1)} \herm{(\tilde\bE^{(t+1)})}\hat\bmp^{(t)}}{\|\tilde\bE^{(t+1)} \herm{(\tilde\bE^{(t+1)})}\hat\bmp^{(t)}\|^2_2},
\end{align}
where we initialize the power method with the subspace estimate $\hat\bmp^{(t)}$  from the previous iteration.

\subsection{The MAED Algorithm}
We now have all the building blocks for MAED, which is summarized in \fref{alg:maed}. Its only input is the receive matrix $\bY$. 
MAED is initialized with $\tilde\bS_D^{(0)}=\mathbf{0}_{U\times D}$, $\tilde\bP^{(0)} = \bI_B,$ 
and $\tau^{(0)}=\tau_0=0.1$, and runs for a fixed number of $t_{\max}$~iterations.

\begin{figure*}[tp]
\centering
\!\!
\subfigure[strong barrage jammer (J1)]{
\includegraphics[height=4cm]{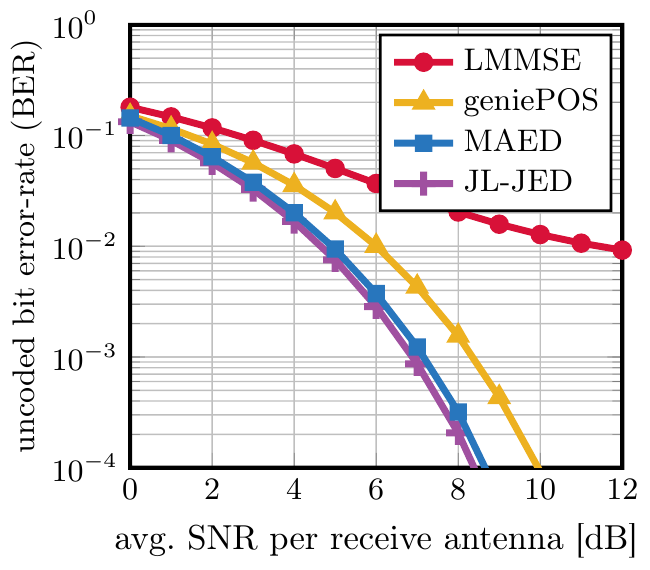}
\label{fig:strong:static}
}\!\!\!
\subfigure[strong pilot jammer (J2)]{
\includegraphics[height=4cm]{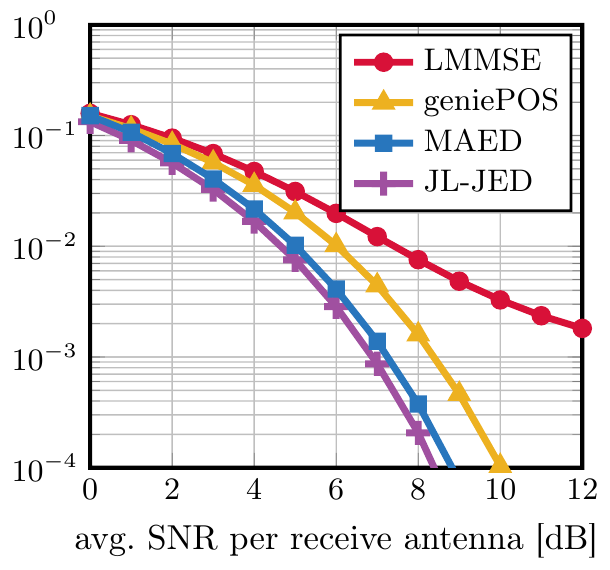}
\label{fig:strong:pilot}
}\!\!\!
\subfigure[strong data jammer (J3)]{
\includegraphics[height=4cm]{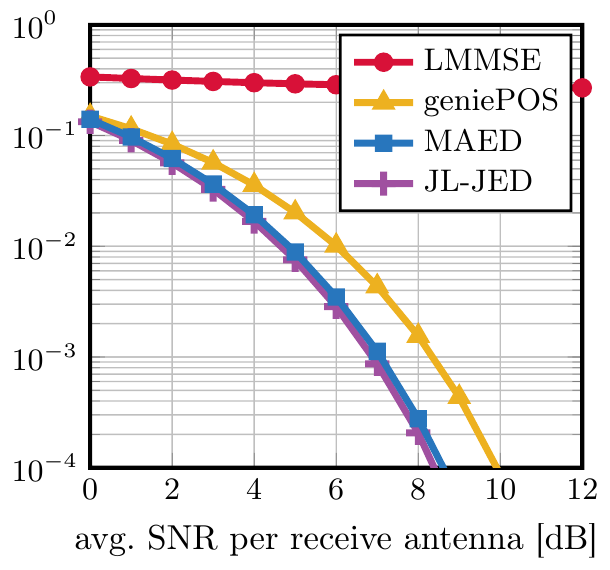}
\label{fig:strong:data}
}\!\!\!
\subfigure[strong sparse jammer (J4)]{
\includegraphics[height=4cm]{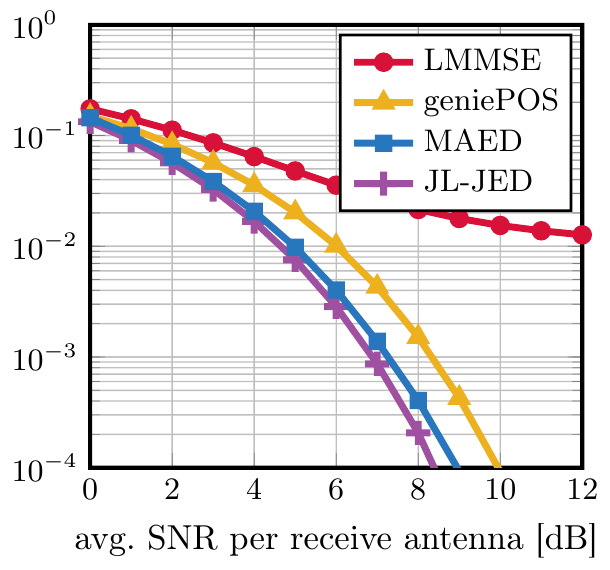}
\label{fig:strong:burst}
}\!\!
\caption{Uncoded bit error-rate (BER) for different detectors in the presence of a \emph{strong} jammer.
The jammer transmits Gaussian symbols either (a) during the entire coherence interval,
(b) during the pilot phase only, (c) during the data phase only, or (d) in random unit-symbol bursts 
with a duty cycle of $\alpha=20\%$. The jammer receive energy over the whole coherence interval is  $\rE=25$\,dB higher than that of the average UE. 
}
\label{fig:strong_jammers}
\end{figure*}

\begin{figure*}[tp]
\centering
\!\!
\subfigure[weak barrage jammer (J1)]{
\includegraphics[height=4cm]{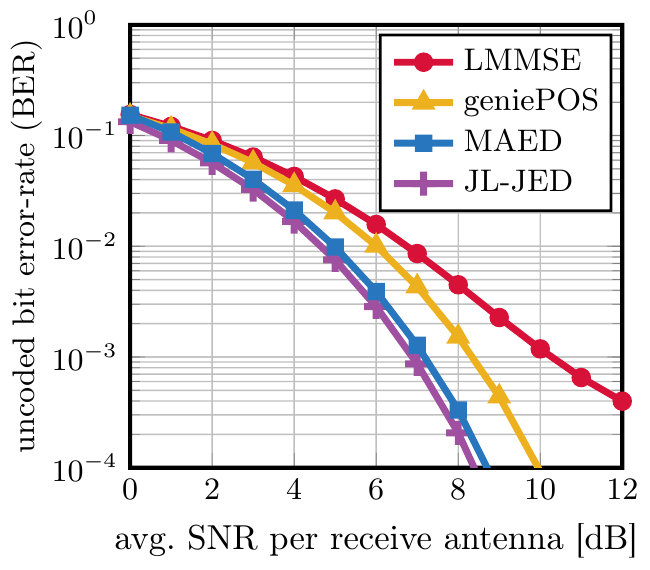}
\label{fig:weak:static}
}\!\!\!
\subfigure[weak pilot jammer (J2)]{
\includegraphics[height=4cm]{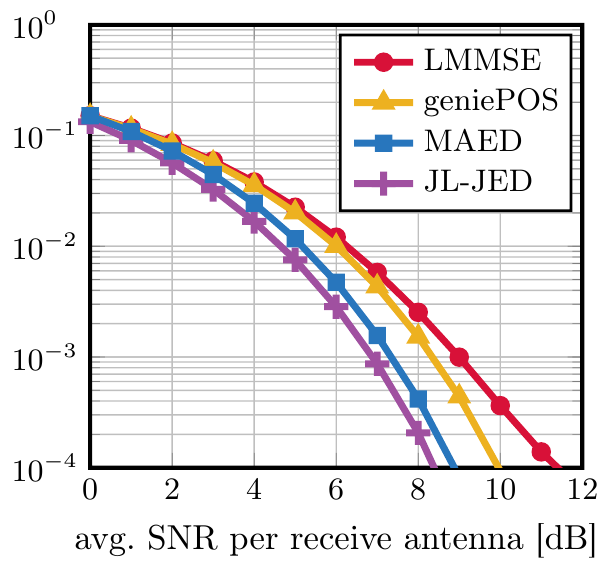}
\label{fig:weak:pilot}
}\!\!\!
\subfigure[weak data jammer (J3)]{
\includegraphics[height=4cm]{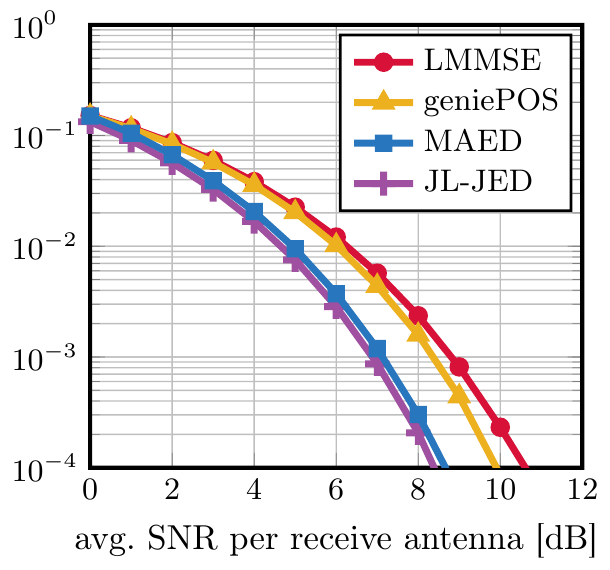}
\label{fig:weak:data}
}\!\!\!
\subfigure[weak sparse jammer (J4)]{
\includegraphics[height=4cm]{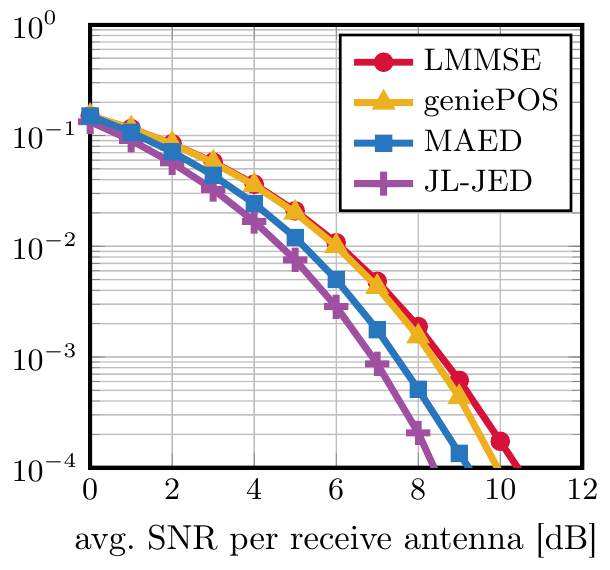}
\label{fig:weak:burst}
}\!\!
\caption{Uncoded bit error-rate (BER) for different detectors in the presence of a \emph{weak} jammer.
The jammer transmits QPSK symbols either (a) during the entire coherence interval,
(b) during the pilot phase only, (c) during the data phase only, or (d) in random unit-symbol bursts 
with a duty cycle of $\alpha=20\%$. The jammer receive power during the jamming phase is as high as that of the average UE ($\rP=0$\,dB).  
}
\label{fig:weak_jammers}
\end{figure*}

\section{Simulation Results}

\subsection{Simulation Setup} \label{sec:setup}
We simulate a massive MU-MIMO system as described in Section~\ref{sec:setup} with $B=128$ BS antennas, 
$U=32$ single-antenna UEs, and with one single-antenna jammer. 
The UEs transmit for $K=96$ time slots.
The first $T=32$ time  slots are used to transmit orthogonal pilots $\bS_T$ in the form of a
 $32\times32$ Hadamard matrix (scaled to symbol energy $\Es$). 
The remaining $D=64$ time slots are used to transmit QPSK payload data (also with symbol energy $\Es$).
The channels of the UEs and the jammer are modelled as i.i.d. Rayleigh fading. 
We define the average receive signal-to-noise ratio (SNR) as follows: 
\begin{align}
\textit{SNR} \define \frac{\Ex{\bS}{\|\bH\bS\|_F^2}}{\Ex{\bN}{\|\bN\|_F^2}}.
\end{align}
In our evaluation, we consider four different types of jammers:  
(J1) barrage jammers that transmit i.i.d.\ jamming symbols during the entire coherence interval,
(J2) pilot jammers that transmit i.i.d.\ jamming symbols during the pilot phase but do not jam the data phase, 
(J3) data jammers that transmit i.i.d.\ jamming symbols during the data phase but do not jam the pilot phase, 
(J4) sparse jammers that transmit i.i.d.\ jamming symbols during some fraction $\alpha$ of randomly selected bursts of unit length (i.e., one time slot), but do not jam the remaining time slots.
The jamming symbols are either circularly symmetric complex Gaussian or selected uniformly at random from the QPSK constellation.
Unless stated otherwise, the jamming symbols are also independent of the UE transmit symbols $\bS$.
We quantify the strength of the jammer's interference relative to the strength of the average UE, either as the ratio 
between total receive \textit{energy}
\begin{align}
\rE \define \frac{\Ex{\bsj}{\|\Hj\bsj\|_2^2}}{\frac1U\Ex{\bS}{\|\bH\bS\|_2^2}},
\end{align}
or as the ratio between receive \textit{power during those phases that the jammer is jamming}
\begin{align}
	\rP = \frac{\rE}{\lambda},
\end{align}
where $\lambda$ is the jammer's duty cycle and equals $1,\frac{T}{K},\frac{D}{K}$, or~$\alpha$
for barrage, pilot, data, or sparse jammers, respectively.

\begin{algorithm}[tp]
\caption{MAED}
\label{alg:maed}
\begin{algorithmic}[1]
\setstretch{1.05}
\vspace{1mm}
\State {\bfseries input:} $\bY$
\State \textbf{initialize:} $\tilde\bS^{(0)} = \left[\bS_T, \mathbf{0}_{U\times D} \right],\,
		\tilde\bP^{(0)} = \bI_B,\, \tau^{(0)} = \tau_0$
\For{$t=0$ {\bfseries to} $t_{\max}-1$}
\State $\nabla f(\tilde\bS^{(t)}) = \herm{\big(\tilde\bS^{(t)}{}^\dagger\big)} \herm\bY\tilde\bP^{(t)}\bY(\bI_K - \tilde\bS^{(t)}{}^\dagger \tilde\bS^{(t)})$
\State $\tilde\bS^{(t+1)} = \proxg\big(\tilde\bS^{(t)} + \tau^{(t)}\nabla f(\tilde\bS^{(t)})\big)$ \hfill(cf.\,\eqref{eq:proxg})
\State $\tilde\bE^{(t+1)} = \bY(\bI_K - \tilde\bS^{(t+1)}{}^\dagger \tilde\bS^{(t+1)})$
\If{$t=0$}
\State $\hat\bmp^{(t+1)} = \bmv_1(\tilde\bE^{(t+1)} \tilde{\bE}^{(t+1)}{}^\text{H})/\|\bmv_1(\tilde\bE^{(t+1)} \tilde{\bE}^{(t+1)}{}^\text{H})\|_2$
\Else
	\State $\hat\bmp^{(t+1)} = \tilde\bE^{(t+1)} \tilde{\bE}^{(t+1)}{}^\text{H}\, \hat\bmp^{(t)}/\|\tilde\bE^{(t+1)} \tilde{\bE}^{(t+1)}{}^\text{H}\, \hat\bmp^{(t)}\|_2$
\EndIf
	\State $\tilde\bP^{(t+1)} = \bI_B - \hat\bmp^{(t+1)}\hat\bmp^{(t+1)}{}^\text{H}$
	\State $\tau^{(t+1)} = \text{Barzilai-Borwein}(\tau^{(t)},\tilde\bS^{(t)},\tilde\bS^{(t+1)}, \dots
	\newline ~\hspace{4.25cm}\!\nabla f(\tilde\bS^{(t)}), \!\nabla f(\tilde\bS^{(t+1)}))$ \hfill\cite{goldstein16a}
	\EndFor
	\State \textbf{output:} $\tilde\bS^{(t_{\max})}_{[T+1:K]}$
\end{algorithmic}
\end{algorithm}

\subsection{Performance Baselines} \label{sec:baseline}
Unless stated otherwise, we run MAED with $t_{\max}=30$ iterations and compare it  with the following baseline methods: 
The first baseline called ``LMMSE'' does not mitigate the jammer in any way and separately performs least-squares channel estimation
and LMMSE data detection.
The second baseline called ``geniePOS'' represents a jammer-robust variant of LMMSE that is furnished with ground-truth knowledge of the jammer channel $\Hj$ and projects the receive
signals~$\bY$ onto the orthogonal complement of $\textit{span}(\Hj)$ as in \eqref{eq:pos}. 
The method then separately performs least-squares channel estimation and LMMSE data detection in this projected subspace.
The third baseline called ``JL-JED'' serves as a performance upper bound and operates in a jammerless but otherwise equivalent scenario. JL-JED performs joint channel estimation 
and data detection by approximately solving~\fref{eq:ml_jed} (with $\setS$ relaxed to its convex hull $\setC$) using the same FBS 
procedure as MAED (cf.~\fref{sec:fbs}), except that it misses the projection~$\tilde\bP$.

\subsection{Mitigation of Strong Gaussian Jammers}
We first investigate the ability of MAED to mitigate strong jamming attacks. 
For this, we simulated Gaussian  jammers with \mbox{$\rE=25$\,dB} of all four jamming types introduced in Section~\ref{sec:setup} 
and evaluated the performance of MAED compared to the baselines of Section~\ref{sec:baseline} (\fref{fig:strong_jammers}).
We point out that the performance of geniePOS and JL-JED is independent of the considered jammer: geniePOS uses the genie-provided
jammer channel to null the jammer perfectly, and JL-JED operates on a jammerless system from the beginning.
Unsurprisingly, the jammer-oblivious LMMSE baseline performs significantly worse than the jammer-robust geniePOS baseline under all attack scenarios.
MAED succeeds in mitigating all four jamming attacks with very high effectiveness, even outperforming the genie-assisted geniePOS method by a
considerable margin.\footnote{The potential for MAED to outperform geniePOS is a consequence of the superiority of JED over 
separating channel estimation from data detection.}
The efficacy of MAED is further reflected in the fact that its BER approaches the BER of the jammerless reference baseline JL-JED 
to within $1$\,dB in all considered scenarios.

\subsection{Mitigation of Weak QPSK Jammers}
We now turn to the analysis of more restrained jamming attacks in which the jammer transmits QPSK symbols 
with relative power $\rP=0$\,dB during its on-phase (to pass itself off as just another UE, for instance \cite{vinogradova16a}).
For now, we still make the assumption that the jamming symbols are independent of the UE transmit matrix $\bS$. 
(We will consider an alternative scenario in~\fref{sec:smart}.)
Simulation results for all four jammer types are shown in~\fref{fig:weak_jammers}. 
The baseline performance of geniePOS and \mbox{JL-JED} are again independent of the jammer and mirror the curves of~\fref{fig:strong_jammers}.
Because of the weaker jamming attacks, the jammer-oblivious LMMSE baseline performs much closer to the jammer-resistant geniePOS baseline.
MAED again mitigates all attack types successfully, outperforming geniePOS and approaching the JL-JED baseline to within $1$\,dB.

Comparing~\fref{fig:weak_jammers} with~\fref{fig:strong_jammers} reveals an interesting phenomenon.
MAED achieves better absolute performance under strong jamming attacks than under weak ones, even if the difference is subtle.
The reason for this behavior is the following: MAED searches for the jamming subspace by looking for a dominant dimension 
of the iterative residual error $\tilde\bE^{(t)}$, see~\fref{eq:rayleigh}. If the received jamming energy is small compared to the 
received signal energy, then 
it becomes harder to distinguish the residual errors due to the jammer's impact from those that are caused by the 
errors in the channel and transmit matrix estimates $\tilde\bH_\bP^{(t)}$ and $\tilde\bS^{(t)}$.

\subsection{What if No Jammer Is Present?}~\label{sec:no_jammer}
\begin{figure}[tp]
\centering
\!\!
\subfigure[no jammer]{
\includegraphics[height=3.85cm]{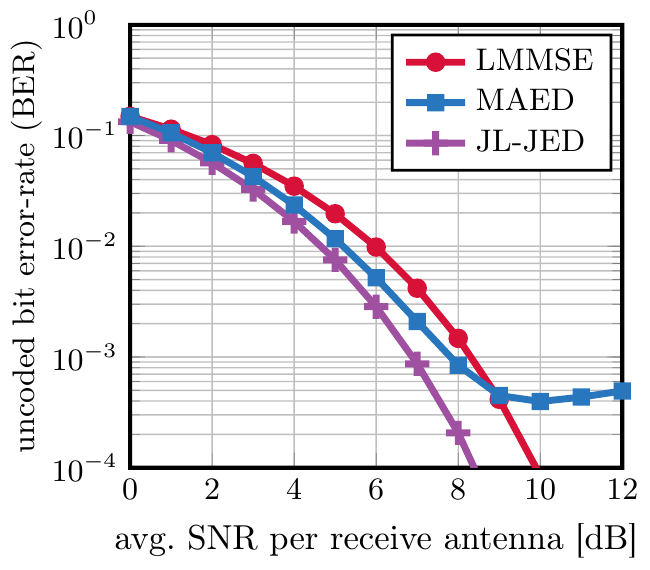}
\label{fig:nojam:no}
}\!\!\!
\subfigure[speed of convergence]{
\includegraphics[height=3.85cm]{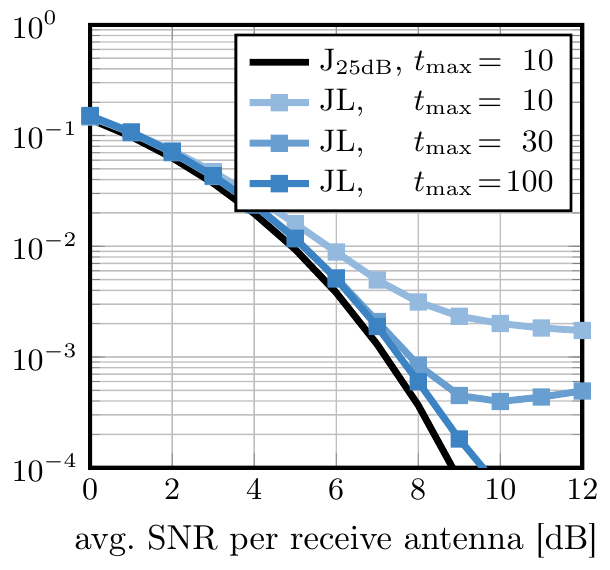}
\label{fig:nojam:speed}
}\!\!
\caption{Uncoded bit error-rate (a) for different detectors in absence of a jammer and (b) for MAED with different number $t_{\max}$ of iterations 
in absence of a jammer (JL), as well as in the presence of a barrage jammer that transmits Gaussian symbols at $\rE=25$\,dB higher energy than the average UE~(J$_{25\text{dB}}$).
}
\label{fig:no_jammer}
\end{figure}

This observation leads to the question of how MAED performs if no jammer is present, or---equivalently---if a jammer does not transmit for a given
coherence interval.~\fref{fig:nojam:no} shows simulation results for this scenario.
MAED still outperforms the LMMSE baseline at low SNR, but shows an error floor at high SNR.
This error floor is caused by the slower convergence of MAED with (infinitely) weak jammers.
\fref{fig:nojam:speed}~shows the jammerless performance of MAED for different numbers~$t_{\max}$ of algorithm iterations.
For $t_{\max}=100$ iterations, MAED essentially achieves the excellent performance that it has in combination with strong jammers.
For $t_{\max}=10$ iterations, however, MAED exhibits an error floor as high as $0.2\%$.
In contrast, in the presence of a $\rE=25$\,dB strong barrage Gaussian jammer, MAED requires no more than $t_{\max}=10$ iterations
for optimal~performance.

The slow convergence in the absence of a jammer can be explained by the fact that, in every iteration, the strongest dimension of the residual 
error matrix $\tilde\bE^{(t)}$ is mistakenly attributed to a hypothesized jammer instead of 
to the residual errors in the channel and transmit matrix estimates $\tilde\bH_\bP^{(t)}$~and~$\tilde\bS^{(t)}$. 
This recurring misattribution prevents fast convergence. 
Nonetheless, while MAED was conceived for jammer mitigation, it shows robust performance even in the absence of jamming.

\subsection{What Happens with a \textit{Truly} Smart Jammer?} \label{sec:smart}

\begin{figure}[tp]
\centering
\!\!
\subfigure[single pilot sequence]{
\includegraphics[height=3.85cm]{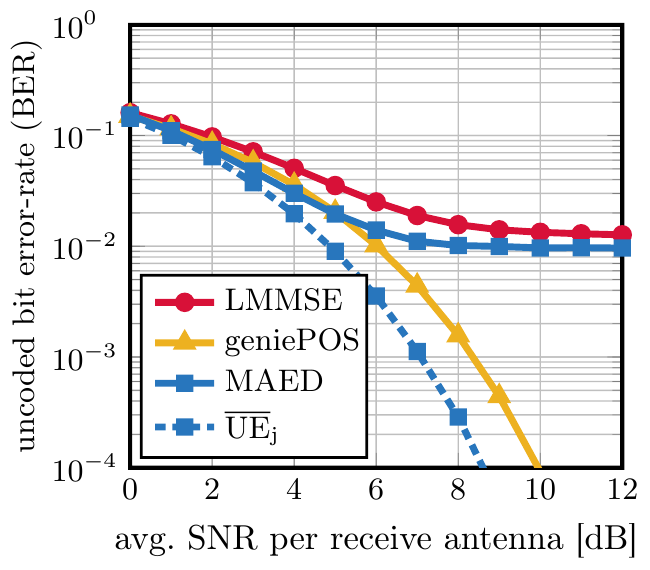}
\label{fig:smart:single}
}\!\!\!
\subfigure[combination of pilot sequences]{
\includegraphics[height=3.85cm]{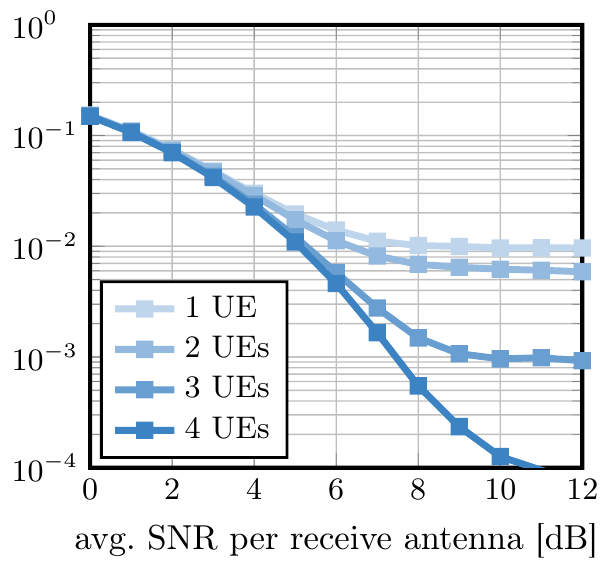}
\label{fig:smart:multi}
}\!\!
\caption{Uncoded bit error rate (BER) for a smart jammer that (a) transmits the pilot
sequence of UE$_\text{j}$ or (b) transmits the average of multiple UE pilot sequences.
The jammer transmits at $\rP=25$\,dB higher power than the average UE.
The $\overline{\text{UE}}_\text{j}$ curve depicts the BER averaged over all UEs except 
$\text{UE}_\text{j}$.
}
\label{fig:smart_jammer}
\end{figure}

Finally, we turn to a scenario in which the jammer knows the UE pilot sequences and attacks a specific UE
by transmitting that UE's pilot sequence during the pilot phase (at $\rP=25$\,dB higher power). 
The jammer does not transmit during the data phase. 
\fref{fig:smart_jammer} shows simulation results for this scenario. 
The geniePOS baseline nulls the jammer perfectly using its ground-truth knowledge. Thus, its performance remains unaffected regardless of the jammer.
In contrast, MAED exhibits an error floor as high as $1\%$, only marginally outperforming the LMMSE baseline.
Excluding the attacked UE and evaluating the BER among the remaining $31$ UEs (labeled $\overline{\text{UE}}_\text{j}$ in \fref{fig:smart:single}) 
reveals that the decoding errors are focused entirely on the attacked UE, and that the BER among the 
remaining UEs appears to be unaffected by the jammer. 
This experiment shows that MAED cannot identify the jammer's subspace if the jammer passes itself off as a UE by transmitting that UE's pilot sequence. 
It is not clear, however, whether such a jammer could be distinguished 
from a legitimate UE, even in principle. 
One way to prevent smart jammers from utilizing such impersonation attacks would be to use encrypted pilot sequences~\cite{basciftci2015securing}.

Finally, \fref{fig:smart:multi} shows the performance of \mbox{MAED} (over all 32 UEs) when the jammer 
transmits the average of multiple pilot sequences during the pilot phase (and refrains from transmitting during the data phase).
Evidently, a jammer that targets multiple UEs quickly enables MAED to locate the jammer's subspace and mitigate the jammer effectively.


\section{Conclusions}

We have proposed MAED in order to mitigate  smart jamming attacks on the uplink of massive
MU-MIMO systems.
In contrast to existing mitigation methods, MAED does not rely on jamming activity 
during any particular epoch for successful jammer mitigation.
Instead, our method exploits the fact that the jammer's subspace remains constant
within a coherence interval.
To this end, MAED uses a novel problem formulation that 
combines jammer estimation and mitigation, channel estimation, and data 
detection. The resulting optimization problem is approximately solved using 
an efficient iterative algorithm. 
Without requiring any \textit{a priori} knowledge,
MAED is able to effectively mitigate a wide range of jamming attacks.
In particular, MAED succeeds in mitigating attack types like data jamming and sparse jamming,
for which---to the best of our knowledge---no mitigation methods have existed so far.

\balance


\end{document}